# Magnetically tunable *rf* wave absorption in polycrystalline $La_{0.67}Ba_{0.33}MnO_3$

V. B. Naik and R. Mahendiran[1]

Department of Physics and NUS Nanoscience & Nanotechnology Initiative

(NUSNNI), Faculty of Science, National University of Singapore,

2 Science Drive 3, Singapore -117542

**Abstract**

We investigated temperature and magnetic field dependent radio-frequency electromagnetic absorption in $La_{0.67}Ba_{0.33}MnO_3$ by monitoring changes in resonance frequency ($f_r$) and current ($I$) through a LC resonant circuit powered by an integrated chip oscillator. The ferromagnetic to paramagnetic transition at $T_c$ in zero external magnetic field is accompanied by an abrupt increase in $f_r$ and $I$ and they are tunable by small external magnetic field. We observed fractional changes as much as 46% in $\Delta f_r/f_r$ and 23% in $\Delta I/I$ around $T_c$ under $\mu_0 H$ = 0.1 T that can be exploited for low magnetic field sensors and other applications.

[1] Corresponding author – phyrm@nus.edu.sg

The perovskite manganite such as $La_{0.67}Ba_{0.33}MnO_3$ have been extensively investigated over the past one decade due the exotic physics involved in colossal negative magnetoresistance shown by them and envisaged applications as magnetic field sensors.[1] For applications as low-magnetic field sensors, a large magnetoresistance, typically $\Delta R/R$ = 10-40 % in a relatively low magnetic field strength, $H$ = 10-100 Gauss is preferable. There are two major contributions to the magnetoresistance (MR) in manganites, one is the intrinsic magnetoresistance arising from the change in the magnetization within a grain or a single domain and another is the extrinsic magnetoresistance due to tunneling of spin-polarized electrons between ferromagnetic grain through semiconducting grain boundaries.[2] Although the intrinsic MR can be nearly 100 % in higher magnetic fields $\mu_0 H \approx$ 3- 6 T around the Curie temperature ($T_c$), it is negligible far below and above the $T_c$. On the other hand, the extrinsic magnetoresistance can reach 20- 40 % in $\mu_0 H$ = 0.3 - 0.5 T at $T$ = 10 K, but becomes only a few percent at room temperature. In this context, observations of a large $\Delta R/R \approx$ 40-80 % in a small field of $H$ = 600 G in microwave frequency ($f$ = 9 -11 GHz)[3, 4, 5] and also in the radio frequency ($f$ = 1- 5 MHz)[6, 7, 8, 9, 10] are quite interesting.

There is another interesting low-field effect called giant magnetoabsorption, i.e., a large change in *rf* electromagnetic wave absorption under a magnetic field that remains less explored. Beletsev *et al.*[6] studied *rf* wave absorption in thin plate of $La_{0.67}Sr_{0.33}MnO_3$ in zero and external magnetic fields using a transmitter–receiver coil technique. Frank J Owen[11] showed that the resonance frequency of tunnel diode powered LC tank circuit with inductor loaded with $La_{0.7}Sr_{0.3}MnO_3$ sample changes as large as $\Delta f_r/f_r$ = -53 % in $\mu_0 H$ = 0.2 T at $f_r$ = 350 kHz around the Curie temperature. Here $\Delta f_r = f_r(T,H)-f_r(T,0)$ is the difference between the resonance frequency of the tank circuit at a given temperature with and without magnetic field ($\mu_0 H$). The large fractional shift in the resonance frequency was attributed to changes in the magnetic penetration depth ($\Delta\delta/\delta \propto -\Delta f_r/f_r$). It was also found that $\Delta f_r/f_r$ is large for certain range of frequencies and decreases at higher frequencies. Srikanth et al.[12] found a much smaller effect ($\Delta f_r/f_r$ = 4% at $\mu_0 H$ = 2 T and $T$ = 100 K) in $Nd_{0.7}Ba_{0.3}MnO_3$. Recently Sarangi et al.[13] observed that rf current through the resonance circuit containing $La_{0.7}Sr_{0.3}MnO_3$ sample changes with the applied magnetic field and temperature, and attributed their observation to the magnetoimpedance effect. A measurement of both resonance frequency and current in the tank circuit will be helpful to understand the nature of electromagnetic absorption in these materials. In this article, we report simultaneous measurements of changes in the resonance frequency and the rf current through the LC tank circuit having $La_{0.67}Ba_{0.63}MnO_3$ sample inside the inductor.

Polycrystalline $La_{0.67}Ba_{033}MnO_3$ was prepared by the standard solid state route and characterized by X-day diffraction, magnetic susceptibility and dc resistivity. The paramagnetic to ferromagnetic transition in our sample as determined by the low frequency ($f$ = 100 Hz) ac susceptibility is $T_c$ = 325 K. The schematic diagram of our experimental set up for the rf power absorption is shown in figure 1. The IC 74LS04 is a TTL bipolar Hex inverter which can be used in high frequency oscillator circuit since the typical propagation delay of each NOT gate is ~15 ns. A dc bias (+5V) is applied to the pin 14 of the IC 74LS04. The pin numbers '1' and '2' are the input and the output of one of the six NOT gates of the IC, respectively. Suppose the voltage at point 'a' is low, then the output at pin '2' is high that builds up the current in the inductor which makes the capacitor to charge; as soon as the voltage at point 'a' is high due the charging of the capacitor, output at pin '2' becomes low and the capacitor starts discharging. This charging and discharging of capacitor ($C$) driven by the NOT gate of the IC through inductor ($L$) sets the LC tank circuit into resonance at frequency, $f_r$ = 1.3 MHz at room temperature.[14] The values of inductance and capacitance at room temperature are 6 µH and 2000 pF, respectively. The resonance frequency is slightly modified from the theoretical value by cable capacitance and resistance in the circuit. A Yogokawa GS 610 source-measure unit (SMU) was used to supply a stable *dc* bias to the IC as well as to measure the *dc* current ($I$) through the circuit. The amplitude of the *ac* voltage across the inductor was measured and found to be 2 V (rms). The *ac* current in the LC circuit was determined by measuring *ac* voltage across a 10 Ohm resistance connected in series with the coil and found to be 7.2 mA (rms) at room temperature which is lower but close to the *dc* current dawn from the power supply. The ac magnetic field is estimated to be $H_{ac}$ = 0.52 Gauss from the *rf* current measured. The inductance coil loaded with the sample was inserted into a commercial cryostat (PPMS, Quantum Design Inc) using coaxial cables whereas the IC and the capacitor were a part of the room temperature circuit. The shift in resonance ($\Delta f_r$) frequency of the tank circuit was tracked using an Agilent 53131 universal frequency counter. When the sample absorbs power from the *rf* field due to changes in its physical property, the impedance of the coil changes which in turn changes the total current drawn from the dc voltage source. If $I(T_1,H_2)$ and $I_0(T_1,H_1)$ are the currents in the circuit under the magnetic fields $H_2$ and $H_1$, and at a temperature $T_1$, the *rf* power absorbed by the sample at any instant of time can be written as $P = V(I-I_0) = V\Delta I$.

Figure 2(a) shows the temperature dependence of the resonance frequency ($f_r$) and figure 2(b) show the current ($I$) through the circuit for various strengths of the external *dc* bias magnetic fields, $\mu_0 H$ = 0.01 T - 0.1 T applied along the coil axis. We have also shown the temperature dependence of $I$ and $f_r$ for the empty coil for comparison. In the absence of external magnetic field ($\mu_0 H$ = 0 T), the resonance frequency of the tank circuit loaded with the sample initially increases gradually with increasing temperature from 10 K but shows an abrupt increase just below the $T_c$ = 325 K and becomes nearly temperature independent above 325 K. It is interesting to note that $f_r$ increases as much as $\Delta f = f_r$ (330 K)-$f_r$(322 K) = (1.26-0.89) MHz = 370 kHz within an interval of 8 K around $T_c$. A small magnetic field of $\mu_0 H$ = 0.01 T hardly affects $f_r$ below 200 K but an appreciable change occurs close to the $T_c$. However, $f_r$ is enhanced over a wide temperature range below the $T_c$ and a broad minimum occurs around $T$ = 186 K < $T_c$ under $\mu_0 H$ = 0.02 T. The largest change in $f_r$ occurs just below the $T_c$. The broad minimum is no more visible at higher fields. Instead, $f_r$ smoothly increases from 10 K and more rapidly as it approaches $T_c$ at higher fields. For $\mu_0 H$ > 0.1 T, $f_r$ appears to approach the value of the empty coil. The current $I$ in the circuit as like the resonance frequency increases sharply around the $T_c$ in $\mu_0 H$ = 0 T upon warming. Although the abrupt change in the current at the $T_c$ is suppressed, the magnitude of the current at the lowest temperature increases with increasing strength of the *dc* magnetic field. We have also find a broad maximum around 100 K in $\mu_0 H$ = 0 T and 0.01 T that is suppressed with increasing field. We do not have a clear understanding of its origin but we note that it may be connected with the change in the behavior of the current seen in the empty coil around 100 K.

We show the temperature dependence of the fractional change in the resonance frequency ($\Delta f_r/f_r = [f_r(H)-f_r(0)]/f_r(0)$) and power absorption ($\Delta P/P = \Delta I/I$) in figure 3 (a) and 3(b), respectively. The $\Delta f_r/f_r$ at $\mu_0 H = 0.01$ T is nearly zero below 290 K but it increases and exhibits a sharp peak of magnitude $\Delta f_r/f_r = 28$ % at $T_c = 325$ K and drops to zero above 328 K. As $H$ increases, the magnitude of the peak also increases and it broadens on the low temperature side. The $\Delta f_r/f_r$ reaches a maximum value of $\approx 46$ % for $\mu_0 H = 0.1$ T at the $T_c$ but it is still large $\approx 37$ % at 10 K. A similar trend is also seen in $\Delta P/P$. The magnitude of the peak increase from $\Delta P/P \approx 13$ % at $\mu_0 H = 0.01$ T to $\approx 23$ % at $\mu_0 H = 0.1$ T. At the same time, the value of $\Delta P/P$ for $T \ll T_c$ also increases with the field. As the result we observe a maximum at $T_c$ instead of a sharp peak at higher fields.

Figure 4 shows the magnetic field dependence of (a) $f_r$ and (b) $I$, at selected temperatures as the field is swept through $\mu_0 H = +400$ mT $\rightarrow -400$m T $\rightarrow +400$ mT. As can be seen in fig 4 (a), the $f_r$ is independent of $H$ above $T_c$. However, $f_r$ changes from almost saturation for $\mu_0 H \geq \pm 300$ mT to a rapid decrease below $\pm 200$ mT followed by a minimum at the origin for $T < T_c$. The current $I$ also shows a similar trend with a minimum at the origin and almost saturation at higher fields. The effect of orientation of the sample with the magnetic field was studied using a variable field electromagnet at room temperature. Figure 5 (a) shows a comparison of $\Delta P/P$ at T = 300 K in a narrow field interval of -200 mT $\leq \mu_0 H \leq$ 200 mT when the dc magnetic field is parallel and perpendicular to the length of the sample and hence to the coil axis. In the parallel configuration, $H_{ac}$ and $H_{dc}$ are parallel to each other whereas they are transverse to each other in the perpendicular configuration. A similar comparison for $\Delta f_r/f_r$ is shown in fig 5(b). The normalization was done here for $f_r$ and $P$ value at the highest field ($\mu_0 H = 200$ mT).  In the parallel configuration, $\Delta P/P$ shows a plateau between -200 mT $\leq \mu_0 H \leq$ 200 mT followed by increase in the intermediate fields range and saturation at $\mu_0 H = \pm$ 0.2 T. In the perpendicular configuration, the plateau extends to higher fields ($\mu_0 H = \pm 60$ mT) and the saturation is not achieved even at $\mu_0 H = 0.2$ T.  The overall magnitude of the $\Delta P/P$ also decreases from 4 % for parallel to 3 % for perpendicular configuration. The $\Delta f_r/f_r$ shown in figure 5(b) shows a similar non saturation behavior in the perpendicular configuration. The change in $\Delta f_r/f_r$ is larger $\approx$ 25 % for the parallel case than $\approx$ 20 % for the perpendicular case. These differences can be attributed to differences in the internal magnetic fields ($H_{int} = H_{dc}+H_{demag}$) felt by the sample as the demagnetizing fields ($H_{demag} = -DM_{sat}$) are different when the applied dc magnetic field is parallel ($D = 0$) and perpendicular ($D = ½$) to the length of sample.

The resonance frequency of the parallel LC tank circuit is $f_r = \frac{1}{2\pi\sqrt{LC}}$ where $L$ is the inductance of the empty coil and $C$ is the capacitance in the circuit. The resonance frequency, upon insertion of a metallic sample, changes due to two reasons, one is due to the true magnetic response, i.e., a change in the real part $rf$ magnetic permeability $\mu'$ of the sample following the magnetic induction $B_{ac} = \mu H_{ac}$,[15] and the other one is the change in the electromagnetic permeability due to the induced eddy currents which screens the alternating magnetic field from the interior of the sample.[16] The relative change in the resonance frequency due to a small change in the penetration depth can be written as $\frac{\Delta f_r}{f_r} = -2\frac{\Delta \delta}{\delta}$ based on $\delta = \sqrt{(2\rho/\omega\mu)}$. If the inductance of the coil changes by a small value $\Delta L$ ($\ll L$), then the fractional change in the resonance frequency can be approximated as $\frac{\Delta f_r}{f_r} \approx -\frac{1}{2}\frac{\Delta L}{L} = -\frac{1}{2}(\eta\chi' + 2\eta\frac{\Delta\delta}{\delta})$ where $\eta = v_s/v_c(1-D)$ is the filling factor, $v_s$ and $v_c$ are the volumes of the sample and the coil respectively, $D$ is the demagnetization factor, $\chi'$ is the real part of the $rf$ susceptibility of the sample and $\delta = \sqrt{(2\rho/\omega\mu)}$ is the skin depth. In our experiment, $\eta = 0.24$. For a highly insulating sample, the dominant contribution to $\Delta f_r/f_r$ comes mainly from $\chi'$ whereas incomplete penetration of the $ac$ magnetic field due to the induced eddy current causes additional contribution in the metallic samples through the second term $\Delta\delta/\delta$. Thus, the rapid increase in $f_r$ around $T_c$ in zero field can be attributed to the abrupt decrease in the real part of the initial $rf$ susceptibility $or$ permeability ($\mu'$) upon transition from the ferromagnetic to paramagnetic state while warming. As our sample is metallic ($d\rho/dT < 0$), the penetration depth ($\delta$) increases with increasing temperature from its lowest value, 10 K and it is expected to show a minimum at $T_c$ and hence $f_r$ shows a minimum at $T_c$. The application of the external magnetic field decreases the $rf$ permeability and increases the penetration depth which leads to increase in $f_r$. The suppression of spin fluctuations by external magnetic field can further aid a dramatic decrease in $\mu'$ leading to a dramatic change in $\Delta f_r/f_r$ at $T_c$.

The power absorption by the material changes the complex impedance (Z) of the tank circuit. If $R_0$ and $L_0$ represent the resistance and inductance of the empty coil, the impedance of the coil upon insertion of the sample is $Z = R_0 + j\omega\mu L_0$ where $\mu = \mu_0\mu_r$ is the $rf$ permeability of the sample. Since $\mu$ is a complex quantity ($\mu = \mu' - j\mu''$), the impedance of the coil becomes $Z = (R_0 + \omega L_0\eta\mu'') + j\omega L_0\eta\mu'$. Thus, the effective resistance of the coil increases due to the magnetic loss characterized by $\mu''$. The above relation is true for highly insulating materials for which the electromagnetic field completely penetrates the sample and it has to be modified for metals since the eddy current shields the external magnetic field. It was shown earlier[10] that the electrical impedance (Z) of $La_{0.75}Sr_{0.25}MnO_3$ dramatically increases at the $T_c$ even when $\mu_0 H = 0$ T when high frequency current (> 1 MHz) flows through the sample. That was related to increase in the surface impedance of the sample $Z = (1+j)\rho/\delta = \sqrt{(j\omega\mu\rho)}$ where $j = \sqrt{-1}$. A similar situation exists in our experiment where sample is subjected to *rf* magnetic field from the inductor. The real part of the surface impedance $Re(Z) = \sqrt{\omega\rho\sqrt{(\mu'^2 + \mu''^2)} + \mu''}$ depends both on the real and imaginary ($\mu''$) components of the *rf* permeability. The power absorption is related to Re(Z) through the relation $P = \frac{1}{2} H_{ac}^2 Re(Z)$.[17] The Re(Z) increases from low temperature, shows a sharp peak just below the Curie temperature and decreases to a small value in the paramagnetic state. Hence, a sudden increase in the *rf* current through the circuit is expected at $T_c$ as observed. The suppression of $\mu'$ and $\mu''$ by the external magnetic field results in negative magnetoresistance which leads to magnetically tunable *rf* wave absorption in our sample.

In conclusion, we have shown that the ferromagnetic to paramagnetic transition in $La_{0.67}Ba_{0.33}MnO_3$ is accompanied by an abrupt increase in the resonance frequency and current through the ICO powered LC-tank circuit. The observed features are related with the dynamical *rf* permeability and magneto impedance of the sample. The magnetically tunable resonance frequency and power absorption demonstrated here can be exploited for magnetic field sensor and other applications.

R. M. acknowledges the National Research Foundation (Singapore) for supporting this work through the grant NRF-G-CRP 2007-05.

**Figure Captions:**

**Fig. 1** A schematic diagram of the IC based LC oscillator circuit used in the present work. The frequency counter tracks the change in the resonance frequency and the source-measure unit (SMU) is used to provide a constant *dc* bias (+5 V) to the IC as well as to measure the current through the circuit.

**Fig. 2** Temperature dependence of the (a) resonance frequency ($f_r$) and (b) current ($I$) through the circuit for different values of external dc magnetic fields ($H$). The data for empty coil are also included.

**Fig. 3** Temperature dependence of the percentage change in the relative (a) frequency shift ($\Delta f_r/f_r$) and (b) power absorption ($\Delta P/P$) for different values of the external magnetic fields ($H$).

**Fig. 4 (a)** Magnetic field dependence of the (a) resonance frequency ($f_r$) and (b) current ($I$) at selected temperatures.

**Fig. 5** Magnetic field dependence of (a) $\Delta P/P$ and (b) $\Delta f_r/f_r$ at $T$ = 300 K when the coil axis is parallel and perpendicular to the *dc* magnetic field direction. The relative directions of the *dc* and *ac* magnetic fields are shown in the insets.

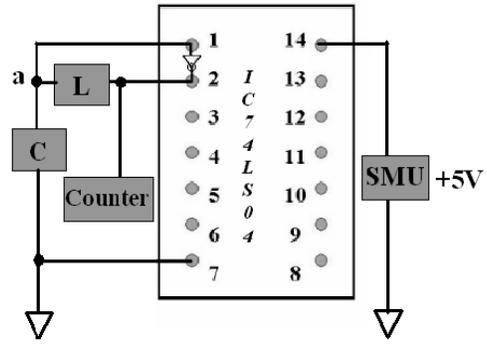

**Fig. 1**
**V. B. Naik** *et al.*

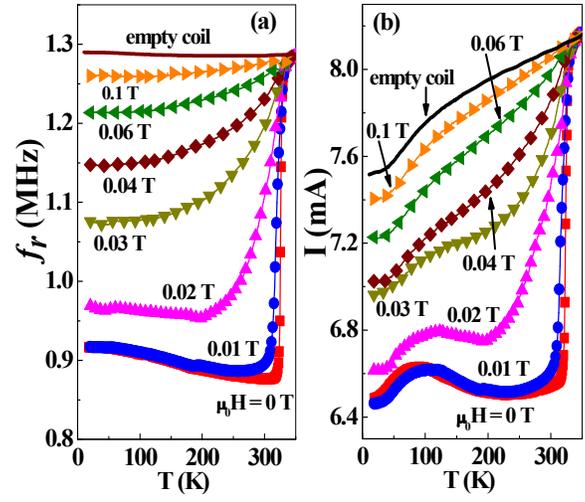

**Fig. 2**
**V. B. Naik *et al.***

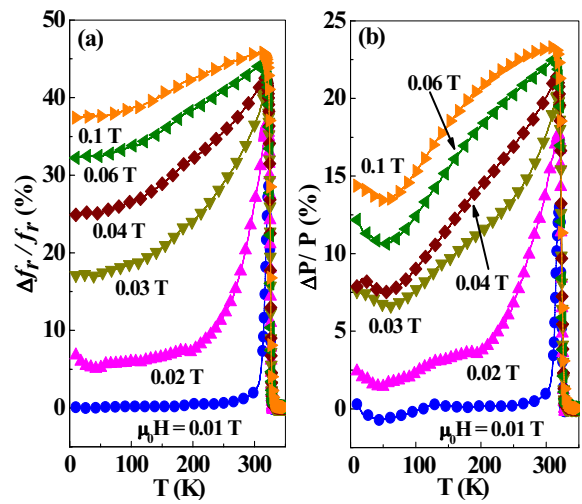

**Fig. 3**
**V. B. Naik** *et al.*

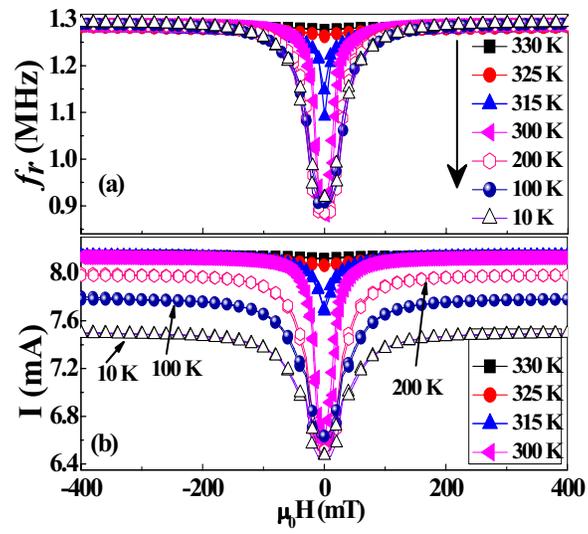

**Fig. 4**
V. B. Naik *et al.*

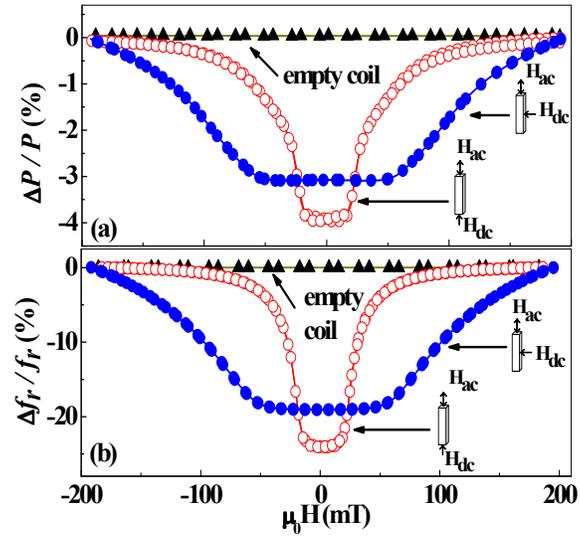

**Fig. 5**
**V. B. Naik** *et al.*